\documentclass[12pt,pdflatex]{iopart}
\usepackage{graphicx}
\usepackage{blackboard}
\def\op#1{\hat{#1}}
\def\ket#1{| #1 \rangle}
\def\bra#1{\langle #1 |}
\def\ip#1#2{\langle #1 \mid #2 \rangle}
\def\ave#1{\langle #1 \rangle}

\def\H{{\cal H}}
\def\L{{\cal L}}
\def\S{{\cal S}}
\def\rmi{i}
\def\trace{\mathop{\rm Tr}}
\newtheorem{definition}{Definition}
\newtheorem{theorem}{Theorem} 
 
\newtheorem{example}{Example}
\newenvironment{proof}{{\sc Proof:} }{\rule{.6em}{0.6em}}
\begin{document}
\bibliographystyle{prsty}
\title{Degrees of controllability for quantum systems and application to atomic systems}
\author{S.~G.\ Schirmer and A.~I.\ Solomon}
\address{Quantum Processes Group, The Open University, 
             Milton Keynes, MK7 6AA, United Kingdom}
\author{J.~V.\ Leahy}
\address{Department of Mathematics and Institute of Theoretical Science, 
             University of Oregon, Eugene, Oregon, 97403, USA}
\eads{\mailto{S.G.Schirmer@open.ac.uk},\mailto{A.I.Solomon@open.ac.uk},\mailto{leahy@math.uoregon.edu}}
\date{\today}
\begin{abstract}
Precise definitions for different degrees of controllability for quantum systems are
given, and necessary and sufficient conditions for each type of controllability are 
discussed.  The results are applied to determine the degree of controllability for 
various atomic systems with degenerate energy levels and transition frequencies.
\end{abstract}
\maketitle

%%%%%%%%%%%%%%%%%%%%%%
\section{Introduction}
%%%%%%%%%%%%%%%%%%%%%%

As control of quantum phenomena has become an increasingly realistic objective due to
both theoretical and technological advances \cite{SCI288p0824}, technical issues such 
as the controllability of quantum systems have received considerable attention recently
\cite{IEEE39CDC1086, IEEE39CDC2803, IEEE39CDC3002, CRAS330p327, PRA63n025403, JPA34p1679, 
MP96p1739, PRA63n063410, qph0106128, CP267p11, CP267p1, qph0110147}.

This has led to the introduction of various definitions of controllability for quantum
systems, such as complete controllability, wavefunction controllability, density matrix 
controllability, observable controllability, etc., as well as a search for necessary and
sufficient conditions to characterize these different degrees of controllability.  Using
general results from Lie group and Lie algebra theory, necessary and sufficient criteria
for various types of controllability have been derived \cite{qph0106128,PRA51p960} and 
for certain types of quantum control systems such as $N$-level ladder systems with dipole
interaction \cite{JPA34p1679, PRA63n063410} or interacting spin systems \cite{qph0106128,
CP267p11, qph0106115}, it has actually been demonstrated that these criteria are satisfied.
Another interesting result is the graph connectivity theorem established in \cite{CP267p1},
which provides sufficient conditions for wavefunction controllability for $N$-level systems, 
if the energy levels are non-degenerate and the transition frequencies are distinct, as well
as generalized versions applicable to other notions of controllability \cite{qph0110147}.

In this paper, we consider the implications of these theoretical results for atomic 
systems with degenerate energy levels and transition frequencies.  The remainder of the
paper is organized as follows.  In section two, we review basic concepts and we give 
precise definitions for various degrees of controllability.  Section three provides a 
concise summary of necessary and sufficient conditions for each type of controllability.
In section four these results are applied to determine the degree of controllability for
atomic transitions between two energy levels with varying degrees of degeneracy.  The 
implications of the results, as well as limits of this approach and directions for 
future work, are discussed in sections five and six, respectively.

%%%%%%%%%%%%%%%%%%%%%%%%
\section{Basic concepts}
%%%%%%%%%%%%%%%%%%%%%%%%

%%%%%%%%%%%%%%%%%%%%%%%%%%%%%%%%%%%%%%%%%%%%%
\subsection{Representation of Quantum States}
%%%%%%%%%%%%%%%%%%%%%%%%%%%%%%%%%%%%%%%%%%%%%

The state of any $N$-level quantum system can be represented by a positive-definite, 
trace-one operator $\op{\rho}$ acting on a Hilbert space.  The representation of the
state is unique under the conditions described in \cite{69Jordan}, chap.\ 5.  This 
so-called \emph{density matrix} always has a discrete spectrum with non-negative 
eigenvalues $w_n$ that sum to one, $\sum_n w_n =1$, and a spectral resolution
\begin{equation} \label{eq:DM}
  \op{\rho} = \sum_{n=1}^N w_n \ket{\Psi_n}\bra{\Psi_n},
\end{equation}
where $\ket{\Psi_n}$ are the eigenstates of $\op{\rho}$ \cite{55N}.  The eigenstates 
corresponding to different eigenvalues are always orthogonal and even if an eigenvalue
occurs with multiplicity greater than one, the $\ket{\Psi_n}$ can always be chosen such
that they form an orthonormal set in the Hilbert space $\H$.  Moreover, by adding states
$\ket{\Psi_n}$ with eigenvalue $w_n=0$ if necessary, we can always assume that 
$\{\ket{\Psi_n}, 1\le n\le N\}$ forms a complete orthonormal set, i.e., a basis for the
Hilbert space $\H$.  The $\bra{\Psi_n}$ are the corresponding dual states defined by
\begin{equation}
  \ip{\Psi_n}{\Psi_m} = \delta_{mn} \qquad \forall m,n.
\end{equation}

\begin{definition}
A density matrix $\op{\rho}$ represents a \emph{pure quantum state} if it has rank one.
In any other case $\op{\rho}$ represents a \emph{mixed quantum state}.
\end{definition}

If $\op{\rho}$ represents a pure state then its eigenvalues counted with multiplicity are
$\{1,0,\ldots,0\}$, i.e., there is one non-zero eigenvalue $w_1=1$ and all other eigenvalues
are zero.  Mixed states must not be confused with superposition states.  For a physical 
system there usually exists a preferred basis consisting of the energy eigenstates of its
internal Hamiltonian.  Superposition states are simply linear combinations of these energy 
eigenstates.  However, they are pure states, as are the energy eigenstates.

\begin{example}
Consider the density matrices
\[
\op{\rho}_1=\left(\begin{array}{cccc} 1 & 0 & 0 & 0 \\
                                      0 & 0 & 0 & 0 \\
                                      0 & 0 & 0 & 0 \\
                                      0 & 0 & 0 & 0 
                   \end{array}\right), \quad
\op{\rho}_2=\left(\begin{array}{cccc} \frac{1}{2} & 0 & 0 & \frac{1}{2} \\
                                      0 & 0 & 0 & 0 \\
                                      0 & 0 & 0 & 0 \\
                                      \frac{1}{2} & 0 & 0 & \frac{1}{2} 
                   \end{array}\right), \quad
 \op{\rho}_3=\left(\begin{array}{cccc} \frac{1}{2} & 0 & 0 & 0 \\
                                       0 & 0 & 0 & 0 \\
                                       0 & 0 & 0 & 0 \\
                                       0 & 0 & 0 & \frac{1}{2} 
                   \end{array}\right).
\]
$\op{\rho}_1$ represents the basis state $\ket{1}$,  while $\op{\rho}_2$ represents the 
superposition state $\ket{\Psi}=\frac{1}{\sqrt{2}}(\ket{1}+\ket{4})\doteq\frac{1}{\sqrt{2}}
(1,0,0,1)^T$ since we have $\op{\rho}_2=\ket{\Psi}\bra{\Psi}$.  However, both density 
matrices have eigenvalues $\{1,0,0,0\}$ and thus represent pure states.  $\op{\rho}_3$, 
on the other hand, represents a true mixed state since we have $\op{\rho}_3=\frac{1}{2}
\ket{1}\bra{1}+\frac{1}{2}\ket{4}\bra{4}$, i.e., its eigenvalues are $\{\frac{1}{2},
\frac{1}{2},0,0\}$ and $\op{\rho}_3$ has rank two.
\end{example}

%%%%%%%%%%%%%%%%%%%%%%%%%%%%%%%%%%%%%%%%%%%%%%%%%%%%%%
\subsection{Kinematical Equivalence Classes of States}
%%%%%%%%%%%%%%%%%%%%%%%%%%%%%%%%%%%%%%%%%%%%%%%%%%%%%%

Conservation laws such as the conservation of energy and probability require the time 
evolution of any (closed) quantum system to be unitary.  Thus, given a pure state 
$\ket{\Psi_0}$, its evolution is determined by $\ket{\Psi(t)}=\op{U}(t)\ket{\Psi_0}$
where $\op{U}(t)$ a unitary operator for all $t$ and $\op{U}(0)=\op{I}$.  Consequently,
a general quantum state represented by a density operator $\op{\rho}_0$ must evolve 
according to $\op{\rho}(t)=\op{U}(t)\op{\rho}_0\op{U}(t)^\dagger$ with $\op{U}(t)$ 
unitary for all times.  This constraint of unitary evolution induces kinematical 
restrictions on the set of target states that are physically admissible from any given
initial state.

\begin{definition}
Two quantum states represented by density matrices $\op{\rho}_0$ and $\op{\rho}_1$ are
\emph{kinematically equivalent} if there exists a unitary operator $\op{U}$ such that 
\begin{equation} \label{eq:KE}
  \op{\rho}_1 = \op{U} \op{\rho}_0 \op{U}^\dagger.
\end{equation}
\end{definition}

\begin{theorem}
Two density matrices $\op{\rho}_0$ and $\op{\rho}_1$ are kinematically equivalent if
and only if they have the same eigenvalues.
\end{theorem}

\begin{proof}
Given two $N\times N$ density matrices that have the same set of eigenvalues, we can 
always find two sets of $N$ orthonormal eigenvectors and a unitary operator $\op{U}\in
U(N)$ that maps one set of eigenvectors onto the other such that $\ket{\Psi_n}=\op{U}
\ket{\phi_n}$ for all $n$.  Thus, we have
\[
  \op{\rho}_1 = \sum_n w_n \ket{\Psi_n}\bra{\Psi_n}
              = \sum_n w_n \op{U}\ket{\phi_n}\bra{\phi_n} \op{U}^\dagger
              = \op{U} \op{\rho}_0\op{U}^\dagger.
\]
Similarly, if (\ref{eq:KE}) holds then $\op{\rho}_1$ and $\op{\rho}_2$ must have the
same eigenvalues since suppose $\op{\rho}_0=\sum_n w_n \ket{\phi_n}\bra{\phi_n}$ then
\[
 \op{\rho}_1 = \op{U}\sum_n w_n \ket{\phi_n}\bra{\phi_n}\op{U}^\dagger
             = \sum_n w_n \op{U}\ket{\phi_n}\bra{\phi_n}\op{U}^\dagger
\]
Since $\op{U}$ is unitary, $\op{U}\ket{\phi_n}$ is an orthonormal basis for $\op{\rho}_1$
and the eigenvalues of $\op{\rho}_1$ are $w_n$.
\end{proof}

%%%%%%%%%%%%%%%%%%%%%%%%%%%%%%%%%%%%%%%%%%%%%%%%%%%%%%
\subsection{Dynamical Lie Groups and Reachable States}
%%%%%%%%%%%%%%%%%%%%%%%%%%%%%%%%%%%%%%%%%%%%%%%%%%%%%%

For any given initial state, only states in the same kinematical equivalence class can
possibly be dynamically reachable.  However, the set of dynamically reachable states may
be a subset of the kinematical equivalence class.  In general, the set of states that 
are dynamically accessible from a given initial state depends on the dynamical Lie group
of the system.  Consider a quantum system with a control-linear Hamiltonian 
\begin{equation} \label{eq:H}
  \op{H}[f_1(t),\ldots,f_M(t)] = \op{H}_0 + \sum_{m=1}^M f_m(t) \op{H}_m,
\end{equation}
where the $f_m$, $1\le m\le M$, are (independent) bounded measurable control functions.  
Since the time-evolution operator $\op{U}(t)$ has to satisfy the Schrodinger equation
\begin{equation} \label{eq:SE}
  \rmi\hbar \frac{d}{d t} \op{U}(t) = \op{H}[f_1(t),\ldots,f_M(t)] \op{U}(t),
\end{equation}
only unitary operators of the form
\begin{equation}
  \op{U}(t)=\exp_+\left\{-\frac{\rmi}{\hbar}\op{H}
                          \left[f_1(t),\ldots,f_M(t)\right]\right\},
\end{equation}
where $\exp_+$ denotes the time-ordered exponential, qualify as evolution operators.  
Using the Magnus expansion of the time-ordered exponential, for instance, it can be 
seen that only unitary operators of the form $\exp(\op{x})$, where $\op{x}$ is an 
element in the dynamical Lie algebra $\L$ generated by the skew-Hermitian operators 
$\rmi\op{H}_0,\ldots,\rmi\op{H}_M$, are dynamically realizable.  These unitary 
operators form the dynamical Lie group $S$ of the system.

For practical purposes, it is often more convenient to consider the related Lie algebra
$\tilde{\L}$, which is generated by the trace-zero parts of the operators $\rmi\op{H}_m$,
\begin{equation}
  \tilde{H}_m = \op{H}_m - \frac{\trace(\op{H}_m)}{N} \op{I}_N,
\end{equation}
for $0\le m\le M$, where $\op{I}_N$ is the identity matrix of dimension $N$.  
$\tilde{\L}$ is always a subalgebra of $su(N)$ since it is generated by trace-zero, 
skew-Hermitian matrices.  Furthermore, if $\trace(\op{H}_m)=0$ for all $m$ then 
$\tilde{\L}=\L$; otherwise we have $\L=\tilde{L}\oplus u(1)$ where $u(1)$ is spanned
by the identity matrix $\op{I}_N$.  

%%%%%%%%%%%%%%%%%%%%%%%%%%%%%%%%%%%%
\section{Degrees of controllability}
%%%%%%%%%%%%%%%%%%%%%%%%%%%%%%%%%%%%

%%%%%%%%%%%%%%%%%%%%%%%%
\subsection{Definitions}
%%%%%%%%%%%%%%%%%%%%%%%%

In this section we give precise definitions for various degrees of controllability for
quantum systems, which are relevant in areas such as quantum-state engineering, quantum
chemistry \cite{CP267p1} and quantum computing \cite{PRA54p1715}.

\begin{definition}
A quantum system is \emph{completely controllable} if any unitary evolution operator
$\op{U}$ is dynamically accessible from the identity $\op{I}$.
\end{definition}

That is, a quantum system is completely controllable if there exists $T>0$, a set of
admissible control functions $(f_1(t),\ldots,f_M(t))$ defined for $0\le t\le T$ and a
corresponding trajectory $\op{U}(t)$ satisfying the Schrodinger equation (\ref{eq:SE})
as well as the boundary conditions $\op{U}(0)=\op{I}$ and $\op{U}(T)=\op{U}$.

\begin{definition}
A quantum system is \emph{density matrix controllable} if for any given initial state 
represented by a density matrix $\op{\rho}_0$, all kinematically equivalent states can
be dynamically reached at some later time $T>0$.
\end{definition}

More precisely, a quantum control system is density-matrix controllable if given any 
two kinematically equivalent states $\op{\rho}_0$ and $\op{\rho}_1$, there exists $T>0$,
a set of admissible control functions $(f_1(t),\ldots,f_M(t))$ defined for $0\le t\le T$
and a corresponding evolution operator $\op{U}(t)$ satisfying the Schrodinger equation 
(\ref{eq:SE}) such that $\op{U}(0)=\op{I}$ and $\op{\rho}_1=\op{U}(T)\op{\rho}_0
\op{U}(T)^\dagger$.

\begin{definition}
A quantum system is \emph{pure-state controllable} if for any given pure initial state 
represented by a wavefunction $\ket{\Psi_0}$, any other pure state $\ket{\Psi_1}$ can 
be dynamically reached at some later time $T$.
\end{definition}

More precisely, a quantum system is pure-state controllable if given any two pure states
$\ket{\Psi_0}$, $\ket{\Psi_1}$ there exists $T>0$, a set of admissible control functions 
$(f_1(t),\ldots,f_M(t))$ defined for $0\le t\le T$ and a corresponding evolution operator
$\op{U}(t)$ satisfying the Schrodinger equation (\ref{eq:SE}) such that $\op{U}(0)
\ket{\Psi_0}=\ket{\Psi_0}$ and $\op{U}(T)\ket{\Psi_0}=\ket{\Psi_1}$.

\begin{definition}
A quantum system is \emph{observable controllable} if any observable represented by a 
Hermitian operator $\op{A}$ on $\H$ can dynamically assume any kinematically admissible
expectation value for any given initial state of the system.
\end{definition}

The kinematically admissible expectation values for any observable $\op{A}$ depend on
the initial state of the system.  Precisely, the expectation value (ensemble average)
of the observable $\op{A}$ is bounded by \cite{PRA58p2684}
\begin{equation}
  \sum_{n=1}^N w_n \lambda_{N+1-n} \le \ave{\op{A}(t)} \le \sum_{n=1}^N w_n \lambda_n,
\end{equation}
where $w_n$ are the eigenvalues of the equivalence class of density operators selected
by the initial state, which are counted with multiplicity and ordered in a non-increasing
sequence, and $\lambda_n$ are the eigenvalues of the operator $\op{A}$, also counted 
with multiplicity and ordered in a non-increasing sequence.  The upper bound is assumed
if the system is in state
\begin{equation}
  \op{\rho}_{+} = \sum_{n=1}^N w_n \ket{\Psi_n}\bra{\Psi_n},
\end{equation}
where $\ket{\Psi_n}$ are the eigenstates of $\op{A}$ satisfying $\op{A}\ket{\Psi_n}=
\lambda_n \ket{\Psi_n}$.  Similarly, the lower bound is assumed if the system is in state
\begin{equation}
  \op{\rho}_{-} = \sum_{n=1}^N w_{N+1-n} \ket{\Psi_n}\bra{\Psi_n}.
\end{equation}
If either $\op{A}$ or $\op{\rho}_0$ has eigenvalues occurring with multiplicity greater
than one then there exists a subspace of states for which the upper and lower bounds are
achieved.  Otherwise, there is a unique state for which these bounds are realized.  Any
intermediate value can be achieved for some state $\op{\rho}_1$, which is kinematically
admissible from the initial state $\op{\rho}_0$.

%%%%%%%%%%%%%%%%%%%%%%%%%%%%%%%%%%%%%%%%%%%%%%%%
\subsection{Necessary and sufficient conditions}
%%%%%%%%%%%%%%%%%%%%%%%%%%%%%%%%%%%%%%%%%%%%%%%%

A necessary and sufficient condition for complete controllability is that the dynamical
Lie group $S$ of the system be $U(N)$.  Noting that $S$ is always a subgroup of $U(N)$,
we see immediately that complete controllability is the strongest possible requirement
of controllability.  Since Eq.\ (\ref{eq:SE}) defines a right-invariant control system
on the compact Lie group $U(N)$, or $SU(N)$ if all the Hamiltonians $\op{H}_m$ have 
zero trace, it follows from theorem 7.1 in \cite{72JS} that a necessary and sufficient 
condition for the dynamical Lie group $S$ to be $U(N)$ is that the dynamical Lie algebra
$\L$ be isomorphic to $u(N)$.  Similarly, $S\simeq SU(N)$ if and only if $\L\simeq 
su(N)$.

As pure states can be represented by complex unit vectors, a sufficient condition for
pure-state controllability of a quantum system of dimension $N$ is that its dynamical
Lie group $S$ act transitively on the unit sphere $S^{2N-1}$ in $\CC^N$.  It is easy 
to see that transitive action on $S^{2N-1}$ is necessary for pure-state controllability
as well.  Using classical results on tranformation groups of spheres \cite{43Montgomery} 
it can be shown that the only subgroups of $U(N)$ that act transitively on the unit 
sphere in $\CC^N$ are $U(N)$ itself, $SU(N)$, and if $N$ is even, $Sp(\frac{N}{2})$ 
as well as $Sp(\frac{N}{2})\times U(1)$ with $U(1)=\{e^{i\phi}\op{I}_N\}$ \cite{qph0106128}.
Thus, a quantum system will be pure-state controllable if and only if its dynamical Lie
group $S$ contains one of the Lie groups above.  In \cite{qph0106128} (Theorem 4) it
is shown that this is possible only if the dynamical Lie algebra $\L$ of the system is
$u(N)$, $su(N)$, or if $N$ even, $sp(\frac{N}{2})$ or $sp(\frac{N}{2})\oplus u(1)$.

A necessary and sufficient condition for density matrix controllability is that the
dynamical Lie group $S$ act transitively on \emph{all} equivalence classes of density
matrices.  Clearly, $S\simeq U(N)$ is sufficient for density matrix controllability.
$SU(N)$ is also sufficient since for any target state $\op{\rho}_1$ that is reachable
from a given initial state $\op{\rho}_0$ via a unitary transformation $\op{U}$, there 
exists an equivalent transformation $\tilde{U}$ in $SU(N)$ such that $\op{\rho}_1=
\tilde{U}\op{\rho}_0\tilde{U}^\dagger$.  Pure-state controllability is a prerequisite
for density matrix controllability since a pure state $\ket{\Psi}$ can be represented
by a density matrix $\ket{\Psi}\bra{\Psi}$.  Thus, only systems whose dynamical Lie
group is $U(N)$, $SU(N)$, $Sp(\frac{N}{2})$ or $Sp(\frac{N}{2})\times U(1)$ qualify for
density matrix controllability  However, we shall see that the latter two are \emph{not}
sufficient for density matrix controllability since there are kinematically admissible
mixed states that cannot be reached from certain initial states.

\begin{example}
Let $N=2\ell$ and assume the dynamical Lie algebra $\tilde{\L}$ generated by the 
trace-zero Hamiltonians $\tilde{H}_m$, $0\le m\le M$, is $sp(\frac{N}{2})$.  Then there
exists a $2\ell\times 2\ell$ matrix $\tilde{J}$, which is unitarily equivalent to
\begin{equation} \label{eq:Jsp}
    \op{J}=\left(\begin{array}{cc} 0 & \op{I}_\ell \\ 
                                  -\op{I}_\ell & 0 
                 \end{array}\right)
\end{equation}
such that
\[
  (\rmi\tilde{H}_m)^T\tilde{J}+\tilde{J}(\rmi\tilde{H}_m) =0, \quad \forall m.
\]
Without loss of generality suppose $\tilde{J}=\op{J}$.  Let $\op{x}$ be a $2\ell\times 
2\ell$ diagonal matrix with diagonal $(-w_1,-w_2,\ldots,-w_\ell,w_1,w_2,\ldots,w_\ell)$ 
and set
\begin{equation} \label{eq:rho0}
  \op{\rho}_0=\frac{1}{2\ell}\op{I}_{2\ell}+\op{x}.
\end{equation}
If $0<w_1<w_2<\ldots<w_\ell<\frac{1}{2\ell}$ then the diagonal elements of $\op{\rho}_0$
are distinct, non-negative and sum to one.  Thus $\op{\rho}_0$ is a density matrix of
rank $N$ with $N=2\ell$ distinct eigenvalues.

$\rmi\op{x}$ is skew-Hermitian and satisfies $(\rmi\op{x})^T\op{J}+\op{J}(\rmi\op{x})=0$.
Thus, $\rmi\op{x}$ is an element in the dynamical Lie algebra $\tilde{\L}$.  Hence, it 
has to remain in $\tilde{\L}$ under the action of the dynamical Lie group $S$.  Thus, 
the state
\begin{equation}
  \op{\rho}_1=\frac{1}{2\ell}\op{I}_{2\ell}+\op{y},
\end{equation}
where $\op{y}$ is a $2\ell\times 2\ell$ diagonal matrix with diagonal $(-w_2,-w_1,\ldots,
-w_\ell,w_1,w_2,\ldots,w_\ell)$, which is kinematically equivalent to $\op{\rho}_0$, is 
not dynamically accessible from $\op{\rho}_0$ since $\rmi\op{y}$ does not satisfy 
$(\rmi\op{y})^T\op{J}+\op{J}(\rmi\op{y})=0$ and is thus not in $\tilde{\L}$.
\end{example}

This example shows that $Sp(\frac{N}{2})$ or $Sp(\frac{N}{2})\times U(1)$ does not act
transitively on all equivalence classes of density matrices; in particular it does not
act transitively on equivalence classes of density matrices of rank $N$ with distinct 
eigenvalues.

Density matrix controllability is clearly sufficient for observable controllability 
for it implies that any state in the same kinematical equivalence class as the initial
state can be reached dynamically.  Therefore, any observable can dynamically assume any 
expectation value (ensemble average) that is kinematically allowed for the equivalence
class of states selected by the initial state.  The previous example also shows that
density matrix controllability is a necessary condition for observable controllability
since $\op{\rho}_1$ is itself an observable that assumes its kinematical upper bound
exactly if the system is in state $\op{\rho}_1$, which is not dynamically accessible
from the initial state $\op{\rho}_0$.  (It is important here that $\op{\rho}_1$ and
$\op{\rho}_0$ have $N$ distinct eigenvalues since this guarantees that there is only 
a single state for which $\op{\rho}_1$ assumes its kinematical upper bound.)

All the above observations can be summarized as follows.
\begin{theorem} \label{thm:Lie:alg}
A necessary and sufficient condition for a quantum system with Hamiltonian (\ref{eq:H})
and dynamical Lie algebra $\L$ ($\tilde{\L}$) to be
\begin{enumerate}
\item \emph{completely controllable}     is $\L \simeq  u(N)$;
\item \emph{density-matrix controllable} is $\tilde{\L}\simeq su(N)$;
\item \emph{observable controllable}     is $\tilde{\L}\simeq su(N)$;
\item \emph{pure-state controllable}     is $\tilde{\L}\simeq su(N)$, or if $N$ is even, 
       $\tilde{\L}\simeq sp(\frac{N}{2})$.
\end{enumerate}
\end{theorem}

This theorem is not only theoretically interestesting.  It can be used in practice to
determine the degree of controllability of a quantum system by computing the dynamical
Lie algebra $\tilde{\L}$ of the system and determining whether it is isomorphic to one 
of the algebras above.  For specific model systems one can often even determine the Lie
algebra as a function of the parameters of the model and apply the previous theorem to
decide if the system is controllable and to what degree \cite{JPA34p1679,PRA63n063410}.  

Notice that density matrix and observable controllability are equivalent.  Complete
controllability is theoretically slightly stronger although the difference is subtle
and can usually be ignored in practice.  To see why this is the case, observe that if
$\trace(\op{H}_m)=0$ for all $m$, i.e., in particular if $\trace(\op{H}_0)=0$ then 
there are certain unitary operators that are not dynamically attainable.  For instance,
if $\L=su(2)$ then the unitary operator 
\[
  \op{U} = \left(\begin{array}{cc} 1 & 0 \\ 0 & e^{i\phi} \end{array}\right)
\]
is not dynamically realizable unless $\phi$ is an integer multiple of $2\pi$.  Hence, 
given $\ket{\Psi_0}=\frac{1}{\sqrt{2}}(\ket{1}+\ket{2})$ then the state $\ket{\Psi_1}=
\frac{1}{\sqrt{2}}(\ket{1}+e^{i\phi}\ket{2})$ is technically \emph{not} reachable from
$\ket{\Psi_0}$.  However, the state $\ket{\tilde{\Psi}_1}=\frac{1}{\sqrt{2}}(e^{-i\phi/2}
\ket{1}+e^{i\phi/2}\ket{2})$, which differs from $\ket{\Psi_1}$ only by an absolute 
phase factor $e^{-i\phi/2}$ \emph{is} dynamically reachable.  Therefore, the trace of 
the operator $\op{H}_0$ is usually not significant for practical purposes.

%%%%%%%%%%%%%%%%%%%%%%%%%%%%%%%%%%%%%%%%%%%
\section{Controllability of atomic systems}
%%%%%%%%%%%%%%%%%%%%%%%%%%%%%%%%%%%%%%%%%%%

\begin{figure*}
\input{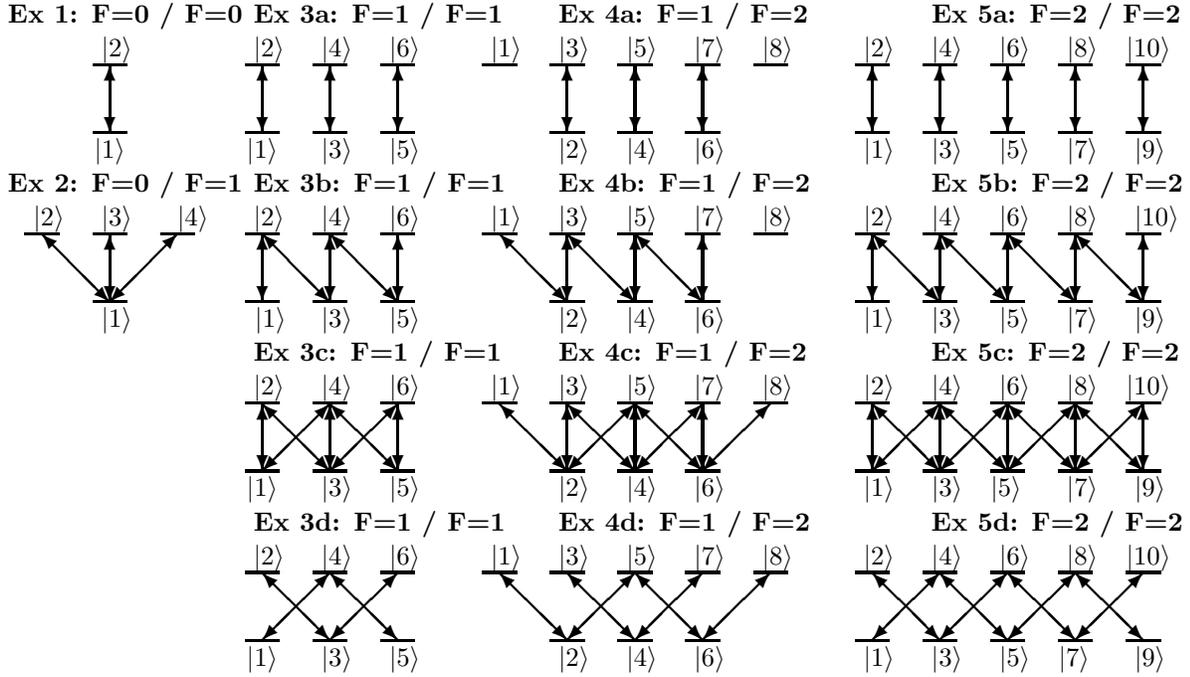}
\caption{Coupling diagrams for various examples. }\label{fig1}
\end{figure*}

We shall now apply the results of the previous sections to determine the degree of 
controllability of various atomic systems.  We consider an atomic system with two main
energy levels, where the number of degenerate sublevels depends on the $F$-value of the
atomic level and is given by the formula $2F+1$.  We can couple magnetic sublevels with
the same quantum number $m$ using a linearly polarized field and sublevels with $\Delta m
=\pm 1$ using left and right circularly polarized fields, respectively.  Fig.~\ref{fig1} 
shows the coupling diagrams for various cases.

%%%%%%%%%%%%%%%%%%%%%%%%%%%%%%%%%%%%%%%%%%%%%%%%
\subsection{Transition between two $F=0$ levels}
%%%%%%%%%%%%%%%%%%%%%%%%%%%%%%%%%%%%%%%%%%%%%%%%

Example 1 in figure \ref{fig1} shows the trivial case of a transition between two 
non-degenerate ($F=0$) levels.  The internal Hamiltonian of this two-state system is
\begin{equation}
  \op{H}_0 = \left(\begin{array}{cc} E_1 & 0 \\ 0 & E_2 \end{array} \right)
\end{equation}
and the interaction Hamiltonian for a linearly polarized control field is 
\begin{equation}
  \op{H}_1 = \left(\begin{array}{cc} 0 & d \\ d & 0 \end{array} \right),
\end{equation}
where $d\neq 0$ is the dipole moment of the transition.  Note that $\rmi\op{H}_0$ and
$\rmi\op{H}_1$ generate $u(2)$ if $E_2\neq -E_1$ and $su(2)$ if $E_2=-E_1$.  Thus, the
system is always density matrix and observable controllable (and completely controllable 
in the former case) using a linearly polarized field.

%%%%%%%%%%%%%%%%%%%%%%%%%%%%%%%%%%%%%%%%%%%%%%%%%%%%%%%%
\subsection{Transitions between $F=0$ and $F=1$ levels}
%%%%%%%%%%%%%%%%%%%%%%%%%%%%%%%%%%%%%%%%%%%%%%%%%%%%%%%%

Example 2 shows a diagram for transitions between sublevels of two atomic levels
with $F=0$ and $F=1$, respectively.  Since the $F=1$ level is three-fold degenerate,
the internal Hamiltonian of this four-state system is given by
\begin{equation}
  \op{H}_0 = \left(\begin{array}{cccc} 
                                 E_1 & 0 & 0 & 0\\
                                 0 & E_2 & 0 & 0\\
                                 0 & 0 & E_2 & 0\\ 
                                 0 & 0 & 0 & E_2 
             \end{array} \right)
\end{equation}
in the standard basis.  Since a linearly polarized field will only drive transitions
between states with the same magnetic quantum number, the interaction Hamiltonian for
a linearly polarized field is
\begin{equation}
  \op{H}_1 = \left(\begin{array}{cccc} 
                                 0 & 0 & d & 0\\ 
                                 0 & 0 & 0 & 0\\
                                 d & 0 & 0 & 0\\ 
                                 0 & 0 & 0 & 0 
             \end{array} \right).
\end{equation}
Similarly, if a left or right circularly polarized field is applied, it will only drive 
transitions between states whose magnetic quantum numbers differ by $\Delta m = \pm 1$, 
respectively.  Thus, the interaction Hamiltonians for a left or right circularly 
polarized field are
\begin{equation}
  \op{H}_2 = \left(\begin{array}{cccc} 
                                 0 & d & 0 & 0\\ 
                                 d & 0 & 0 & 0\\
                                 0 & 0 & 0 & 0\\ 
                                 0 & 0 & 0 & 0 
             \end{array} \right), \quad
  \op{H}_3 = \left(\begin{array}{cccc} 
                                 0 & 0 & 0 & d\\ 
                                 0 & 0 & 0 & 0\\
                                 0 & 0 & 0 & 0\\ 
                                 d & 0 & 0 & 0 
             \end{array} \right).
\end{equation}
It is obvious from the diagram that all three polarizations are required to couple
all the levels and noting that $\rmi\op{H}_0$, $\rmi\op{H}_1$, $\rmi\op{H}_2$ and 
$\rmi\op{H}_3$ generate the Lie algebra $u(4)$ (or $su(4)$ if $E_2=-\frac{1}{3}E_1$)
shows that the system is always density matrix and observable controllable in this
case.

%%%%%%%%%%%%%%%%%%%%%%%%%%%%%%%%%%%%%%%%%%%%%%%%
\subsection{Transition between two $F=1$ levels}
%%%%%%%%%%%%%%%%%%%%%%%%%%%%%%%%%%%%%%%%%%%%%%%%

Example 3 shows four different coupling diagrams for transitions between sublevels
of two atomic levels with $F=1$.  Considering the ordering of the states chosen in
the diagram, the internal Hamiltonian is
\begin{equation} \label{H0:F=1}
  \op{H}_0 = \left(\begin{array}{cc|cc|cc} 
                                E_1 & 0 & 0 & 0 & 0 & 0\\ 
                                0 & E_2 & 0 & 0 & 0 & 0\\\hline 
                                0 & 0 & E_1 & 0 & 0 & 0\\ 
                                0 & 0 & 0 & E_2 & 0 & 0\\\hline 
                                0 & 0 & 0 & 0 & E_1 & 0\\
                                0 & 0 & 0 & 0 & 0 & E_2
             \end{array} \right)
\end{equation}
and the interaction Hamiltonians for linearly, as well as left and right circularly
polarized fields are
\begin{equation} \label{H1:F=1}
  \op{H}_1 = \left(\begin{array}{cc|cc|cc} 
                                 0 & d & 0 & 0 & 0 & 0\\ 
                                 d & 0 & 0 & 0 & 0 & 0\\\hline 
                                 0 & 0 & 0 & d & 0 & 0\\ 
                                 0 & 0 & d & 0 & 0 & 0\\\hline 
                                 0 & 0 & 0 & 0 & 0 & d\\
                                 0 & 0 & 0 & 0 & d & 0
             \end{array} \right), 
\end{equation}
\begin{equation} \label{H2,3:F=1}
  \op{H}_2 = \left(\begin{array}{c|cc|cc|c} 
                                 0 & 0 & 0 & 0 & 0 & 0\\\hline  
                                 0 & 0 & d & 0 & 0 & 0\\
                                 0 & d & 0 & 0 & 0 & 0\\\hline  
                                 0 & 0 & 0 & 0 & d & 0\\
                                 0 & 0 & 0 & d & 0 & 0\\\hline 
                                 0 & 0 & 0 & 0 & 0 & 0
             \end{array} \right), \quad
  \op{H}_3 = \left(\begin{array}{cccccccc} 
                                 0 & 0 & 0 & d & 0 & 0\\ 
                                 0 & 0 & 0 & 0 & 0 & 0\\
                                 0 & 0 & 0 & 0 & 0 & d\\ 
                                 d & 0 & 0 & 0 & 0 & 0\\
                                 0 & 0 & 0 & 0 & 0 & 0\\
                                 0 & 0 & d & 0 & 0 & 0
             \end{array} \right).
\end{equation}

The Lie algebra generated by $\rmi\op{H}_0$ and $\rmi\op{H}_1$ is isomorphic to $u(2)$ 
(or $su(2)$ if $E_2=-E_1$).  Thus, if only a linearly polarized field is used then the
system breaks up into three congruent, non-interacting two-level subsystems (as shown 
in Example 3a) and behaves effectively like a two-state system.

The Lie algebra generated by $\rmi\op{H}_0$, $\rmi\op{H}_1$ and $\rmi\op{H}_2$ is 
isomorphic to $sp(3)\oplus u(1)$ (or $sp(3)$ if $E_2=-E_1$).  Thus, if linearly and 
left circularly polarized fields are used then the system is pure-state controllable
(see Example 3b).  The same holds if linearly and right circularly polarized fields 
are used instead (not shown).

However, even if linearly as well as left \emph{and} right circularly polarized fields
are used (as shown in Example 3c) the system is neither density matrix nor observable
controllable since the Lie algebra generated by $\rmi\op{H}_0$, $\rmi\op{H}_1$, $\rmi
\op{H}_2$ and $\rmi\op{H}_3$ is still isomorphic to $sp(3)\oplus u(1)$ (or $sp(3)$ if 
$E_2=-E_1$) as in the previous case.

Finally, if only left and right circularly polarized fields (see Example 3d) are used 
then the system decomposes into two congruent, non-interacting three-level subsystems 
spanned by the states $\ket{1}$, $\ket{4}$, $\ket{5}$ and $\ket{2}$, $\ket{3}$, $\ket{6}$, 
respectively.  Furthermore, the Lie algebra generated by $\rmi\op{H}_0$, $\rmi\op{H}_2$ 
and $\rmi\op{H}_3$ is $u(3)$, i.e., the system behaves effectively like a completely 
controllable three-state system.

%%%%%%%%%%%%%%%%%%%%%%%%%%%%%%%%%%%%%%%%%%%%%%%%%%%%%%%%
\subsection{Transitions between $F=1$ and $F=2$ levels}
%%%%%%%%%%%%%%%%%%%%%%%%%%%%%%%%%%%%%%%%%%%%%%%%%%%%%%%%

Example 4 shows four different coupling diagrams for transitions between sublevels
of two atomic levels with $F=1$ and $F=2$, respectively.  Given the ordering of 
states chosen in the diagram, the internal Hamiltonian is
\begin{equation}
  \op{H}_0 = \left(\begin{array}{c|cc|cc|cc|c} 
                                 E_2 & 0 & 0 & 0 & 0 & 0 & 0 & 0\\\hline 
                                 0 & E_1 & 0 & 0 & 0 & 0 & 0 & 0\\ 
                                 0 & 0 & E_2 & 0 & 0 & 0 & 0 & 0\\\hline
                                 0 & 0 & 0 & E_1 & 0 & 0 & 0 & 0\\ 
                                 0 & 0 & 0 & 0 & E_2 & 0 & 0 & 0\\\hline
                                 0 & 0 & 0 & 0 & 0 & E_1 & 0 & 0\\
                                 0 & 0 & 0 & 0 & 0 & 0 & E_2 & 0\\\hline
                                 0 & 0 & 0 & 0 & 0 & 0 & 0 & E_2
             \end{array} \right)
\end{equation}
and the interaction Hamiltonians for linearly as well as left and right circularly 
polarized fields are
\begin{equation}
  \op{H}_1 = \left(\begin{array}{c|cc|cc|cc|c} 
                                 0 & 0 & 0 & 0 & 0 & 0 & 0 & 0\\\hline
                                 0 & 0 & d & 0 & 0 & 0 & 0 & 0\\ 
                                 0 & d & 0 & 0 & 0 & 0 & 0 & 0\\\hline
                                 0 & 0 & 0 & 0 & d & 0 & 0 & 0\\ 
                                 0 & 0 & 0 & d & 0 & 0 & 0 & 0\\\hline
                                 0 & 0 & 0 & 0 & 0 & 0 & d & 0\\
                                 0 & 0 & 0 & 0 & 0 & d & 0 & 0\\\hline
                                 0 & 0 & 0 & 0 & 0 & 0 & 0 & 0
             \end{array} \right),
\end{equation}
\begin{equation}
  \op{H}_2 = \left(\begin{array}{cc|cc|cc|cc} 
                                 0 & d & 0 & 0 & 0 & 0 & 0 & 0\\
                                 d & 0 & 0 & 0 & 0 & 0 & 0 & 0\\\hline 
                                 0 & 0 & 0 & d & 0 & 0 & 0 & 0\\ 
                                 0 & 0 & d & 0 & 0 & 0 & 0 & 0\\\hline
                                 0 & 0 & 0 & 0 & 0 & d & 0 & 0\\ 
                                 0 & 0 & 0 & 0 & d & 0 & 0 & 0\\\hline 
                                 0 & 0 & 0 & 0 & 0 & 0 & 0 & 0\\ 
                                 0 & 0 & 0 & 0 & 0 & 0 & 0 & 0 
             \end{array} \right),
\end{equation}
\begin{equation}
  \op{H}_3 = \left(\begin{array}{cccccccc} 
                                 0 & 0 & 0 & 0 & 0 & 0 & 0 & 0\\
                                 0 & 0 & 0 & 0 & d & 0 & 0 & 0\\ 
                                 0 & 0 & 0 & 0 & 0 & 0 & 0 & 0\\
                                 0 & 0 & 0 & 0 & 0 & 0 & d & 0\\ 
                                 d & 0 & 0 & 0 & 0 & 0 & 0 & 0\\
                                 0 & 0 & 0 & 0 & 0 & 0 & 0 & d\\ 
                                 0 & 0 & 0 & d & 0 & 0 & 0 & 0\\
                                 0 & 0 & 0 & 0 & 0 & d & 0 & 0
             \end{array} \right).
\end{equation}
If only a linearly polarized field is used (as shown in Example 4a) then the Lie algebra
of the system is again $u(2)$ (or $su(2)$ if $E_2=-E_1$), i.e., the system behaves effectively
like a two-state system.

If linearly and left circularly polarized fields are used then states $\ket{1}$ through 
$\ket{7}$ form a density matrix and observable controllable subsystem since the Lie 
algebra generated by $\rmi\op{H}_0$, $\rmi\op{H}_1$ and $\rmi\op{H}_2$ is isomorphic to
$u(7)$ (or $su(7)$ in the trace-zero case). 

If linearly as well as left \emph{and} right circularly polarized fields are used (as 
shown in Example 4c) then the system is density matrix and observable controllable since
$\rmi\op{H}_0$, $\rmi\op{H}_1$, $\rmi\op{H}_2$ and $\rmi\op{H}_3$ generate the Lie 
algebra $u(8)$ (or $su(8)$ if $3E_1+5E_2=0$).
  
Finally, if only left and right circularly polarized fields are used then the system 
decomposes into two non-interacting subsystems consisting of states $\ket{3}$, $\ket{4}$,
$\ket{7}$ and states $\ket{1}$, $\ket{2}$, $\ket{5}$, $\ket{6}$, $\ket{8}$, respectively,
as shown in Example 4d.  The system is therefore not controllable.  However, it can be 
verified that each of the \emph{sub}systems is effectively completely controllable. 

%%%%%%%%%%%%%%%%%%%%%%%%%%%%%%%%%%%%%%%%%%%%%%%%
\subsection{Transition between two $F=2$ levels}
%%%%%%%%%%%%%%%%%%%%%%%%%%%%%%%%%%%%%%%%%%%%%%%%

Example 5 shows four different coupling diagrams for transitions between sublevels
of two atomic levels with $F=2$.   Given the ordering of states in the diagram, 
the internal Hamiltonian is
\begin{equation} \label{H0:F=2}
  \op{H}_0 = \left(\begin{array}{cc|cc|cc|cc|cc} 
                                 E_1 & 0 & 0 & 0 & 0 & 0 & 0 & 0 & 0 & 0\\ 
                                 0 & E_2 & 0 & 0 & 0 & 0 & 0 & 0 & 0 & 0\\\hline
                                 0 & 0 & E_1 & 0 & 0 & 0 & 0 & 0 & 0 & 0\\ 
                                 0 & 0 & 0 & E_2 & 0 & 0 & 0 & 0 & 0 & 0\\\hline
                                 0 & 0 & 0 & 0 & E_1 & 0 & 0 & 0 & 0 & 0\\ 
                                 0 & 0 & 0 & 0 & 0 & E_2 & 0 & 0 & 0 & 0\\\hline
                                 0 & 0 & 0 & 0 & 0 & 0 & E_1 & 0 & 0 & 0\\
                                 0 & 0 & 0 & 0 & 0 & 0 & 0 & E_2 & 0 & 0\\\hline
                                 0 & 0 & 0 & 0 & 0 & 0 & 0 & 0 & E_1 & 0\\
                                 0 & 0 & 0 & 0 & 0 & 0 & 0 & 0 & 0 & E_2
             \end{array} \right)
\end{equation}
and the interaction Hamiltonians for linearly as well as left and right circularly 
polarized fields are
\begin{equation} \label{H1:F=2}
  \op{H}_1 = \left(\begin{array}{cc|cc|cc|cc|cc} 
                                 0 & d & 0 & 0 & 0 & 0 & 0 & 0 & 0 & 0\\
                                 d & 0 & 0 & 0 & 0 & 0 & 0 & 0 & 0 & 0\\\hline
                                 0 & 0 & 0 & d & 0 & 0 & 0 & 0 & 0 & 0\\
                                 0 & 0 & d & 0 & 0 & 0 & 0 & 0 & 0 & 0\\\hline
                                 0 & 0 & 0 & 0 & 0 & d & 0 & 0 & 0 & 0\\
                                 0 & 0 & 0 & 0 & d & 0 & 0 & 0 & 0 & 0\\\hline
                                 0 & 0 & 0 & 0 & 0 & 0 & 0 & d & 0 & 0\\
                                 0 & 0 & 0 & 0 & 0 & 0 & d & 0 & 0 & 0\\\hline
                                 0 & 0 & 0 & 0 & 0 & 0 & 0 & 0 & 0 & d\\
                                 0 & 0 & 0 & 0 & 0 & 0 & 0 & 0 & d & 0
             \end{array} \right),
\end{equation}
\begin{equation} \label{H2:F=2}
  \op{H}_2 = \left(\begin{array}{c|cc|cc|cc|cc|c} 
                                 0 & 0 & 0 & 0 & 0 & 0 & 0 & 0 & 0 & 0\\\hline
                                 0 & 0 & d & 0 & 0 & 0 & 0 & 0 & 0 & 0\\
                                 0 & d & 0 & 0 & 0 & 0 & 0 & 0 & 0 & 0\\\hline
                                 0 & 0 & 0 & 0 & d & 0 & 0 & 0 & 0 & 0\\
                                 0 & 0 & 0 & d & 0 & 0 & 0 & 0 & 0 & 0\\\hline
                                 0 & 0 & 0 & 0 & 0 & 0 & d & 0 & 0 & 0\\
                                 0 & 0 & 0 & 0 & 0 & d & 0 & 0 & 0 & 0\\\hline
                                 0 & 0 & 0 & 0 & 0 & 0 & 0 & 0 & d & 0\\
                                 0 & 0 & 0 & 0 & 0 & 0 & 0 & d & 0 & 0\\\hline
                                 0 & 0 & 0 & 0 & 0 & 0 & 0 & 0 & 0 & 0
             \end{array} \right),
\end{equation}
\begin{equation} \label{H3:F=2}
  \op{H}_3 = \left(\begin{array}{cccccccccc} 
                                 0 & 0 & 0 & d & 0 & 0 & 0 & 0 & 0 & 0\\
                                 0 & 0 & 0 & 0 & 0 & 0 & 0 & 0 & 0 & 0\\
                                 0 & 0 & 0 & 0 & 0 & d & 0 & 0 & 0 & 0\\
                                 d & 0 & 0 & 0 & 0 & 0 & 0 & 0 & 0 & 0\\
                                 0 & 0 & 0 & 0 & 0 & 0 & 0 & d & 0 & 0\\
                                 0 & 0 & d & 0 & 0 & 0 & 0 & 0 & 0 & 0\\
                                 0 & 0 & 0 & 0 & 0 & 0 & 0 & 0 & 0 & d\\
                                 0 & 0 & 0 & 0 & 0 & d & 0 & 0 & 0 & 0\\
                                 0 & 0 & 0 & 0 & 0 & 0 & 0 & 0 & 0 & 0\\
                                 0 & 0 & 0 & 0 & 0 & 0 & 0 & d & 0 & 0
             \end{array} \right).
\end{equation}
The Lie algebra generated by $\rmi\op{H}_0$ and $\rmi\op{H}_1$ only is isomorphic to 
$u(2)$ (or $su(2)$ if $E_2=-E_1$).  Thus, if only a linearly polarized field is used 
then the system behaves again effectively like a two-state system.

The Lie algebra generated by $\rmi\op{H}_0$, $\rmi\op{H}_1$ and $\rmi\op{H}_2$ is 
isomorphic to $sp(5)\oplus u(1)$ (or $sp(5)$ if $E_2=-E_1$).  Thus, if linearly and
left circularly polarized fields are used then the system is pure-state controllable
(see Example 5b).  The same is true if linearly and right circularly polarized fields
are used instead (not shown).

However, even if linearly as well as left \emph{and} right circularly polarized fields
are used (as shown in Example 5c) the system is not density matrix controllable as the
Lie algebra generated by $\rmi\op{H}_0$, $\rmi\op{H}_1$, $\rmi\op{H}_2$ and $\rmi\op{H}_3$
is isomorphic to $sp(5)\oplus u(1)$ (or $sp(5)$ if $E_2=-E_1$). 

If only left and right circularly polarized fields are used then the system decomposes
into two congruent, non-interacting five-level subsystems $\ket{1}$, $\ket{4}$, 
$\ket{5}$, $\ket{8}$, $\ket{9}$ and $\ket{2}$, $\ket{3}$, $\ket{6}$, $\ket{7}$, $\ket{10}$
(see Example 5d).  Since the Lie algebra generated by $\rmi\op{H}_0$, $\rmi\op{H}_2$ and
$\rmi\op{H}_3$ is isomorphic to $u(5)$, the system behaves effectively like a completely
controllable five-state system.

%%%%%%%%%%%%%%%%%%%%%%%%%%%%%%%%%%%%%%%%%%%%%%%%%%
\section{Interpretation of results and discussion}
%%%%%%%%%%%%%%%%%%%%%%%%%%%%%%%%%%%%%%%%%%%%%%%%%%

The main results of the analysis in the previous section can be summarized as follows.
\begin{enumerate}
\item A transition between two $F=k$ levels with $k=1,2,3$ is \emph{pure-state controllable} 
      using linearly and \emph{either} left \emph{or} right circularly polarized fields, 
      but it is \emph{not} density matrix or observable controllable even if linearly, 
      left \emph{and} right circularly polarized fields are applied. (The $F=0$ case is
      an \emph{exception} --- a transition between two $F=0$ levels is clearly always
      density matrix and observable controllable using only a linearly polarized field.)
\item A transition between two levels with $F=k$ and $F'=k+1$ for $k=0,1,2$, on the other
      hand, is always \emph{density matrix and observable controllable} using linearly as
      well as left \emph{and} right circularly polarized fields.
\item None of these transitions are controllable using only left and right circularly 
      polarized fields.
\item If only linearly polarized fields are applied then \emph{all} the above transitions 
      behave effectively like two-state systems.
\end{enumerate}

%%%%%%%%%%%%%%%%%%%%%%%%%%%%%%%%%%%%%%%%%%%%%%%%%%
\subsection{Practical implications of the results}
%%%%%%%%%%%%%%%%%%%%%%%%%%%%%%%%%%%%%%%%%%%%%%%%%%

As regards the implications of these results, the last observation is important because
it provides a mathematical \emph{justification} for the quite common practice of treating 
transitions between two energy levels like two-level systems even if the energy levels 
are degenerate.  Indeed, our analysis shows that if only linearly polarized light is used 
(and the sublevel structure is not important for the application at hand) then this 
simplification is mathematically justified.

Another interesting result is that the degree of controllability of a transition between
energy levels with a \emph{different} number of sub-levels is \emph{different} from that
of a transition between two energy levels with the \emph{same} number of sublevels.  In 
the former case, the degree of controllability analysis shows that every unitary operator
in $U(N)$ or $SU(N)$ (where $N$ is the total number of coupled sublevels) is dynamically
realizable, while in the latter case the system is only pure-state controllable no matter
what fields are applied.  To better understand the implications of the latter statement, 
let us consider an example.

\begin{figure}
\begin{center}
\setlength{\unitlength}{2800sp}%
\begingroup\makeatletter\ifx\SetFigFont\undefined%
\gdef\SetFigFont#1#2#3#4#5{%
  \reset@font\fontsize{#1}{#2pt}%
  \fontfamily{#3}\fontseries{#4}\fontshape{#5}%
  \selectfont}%
\fi\endgroup%
\begin{picture}(8880,1155)(293,-361)
\thicklines
\put(1126,689){\line( 1, 0){600}}
\put(1126,239){\line( 1, 0){600}}
\put(2026,689){\line( 1, 0){600}}
\put(2026,239){\line( 1, 0){600}}
\put(2926,689){\line( 1, 0){600}}
\put(2926,239){\line( 1, 0){600}}
\put(2326,239){\vector( 0, 1){450}}
\put(601,614){\makebox(0,0)[b]{\smash{\SetFigFont{10}{12}{\rmdefault}{\mddefault}{\updefault}$P_{3/2}$}}}
\put(601,164){\makebox(0,0)[b]{\smash{\SetFigFont{10}{12}{\rmdefault}{\mddefault}{\updefault}$P_{3/2}$}}}
\put(1276, 14){\makebox(0,0)[lb]{\smash{\SetFigFont{10}{12}{\rmdefault}{\mddefault}{\updefault}$w_{-1}$}}}
\put(2176, 14){\makebox(0,0)[lb]{\smash{\SetFigFont{10}{12}{\rmdefault}{\mddefault}{\updefault}$w_0$}}}
\put(3076, 14){\makebox(0,0)[lb]{\smash{\SetFigFont{10}{12}{\rmdefault}{\mddefault}{\updefault}$w_{+1}$}}}
\put(2326,-361){\makebox(0,0)[b]{\smash{\SetFigFont{10}{12}{\rmdefault}{\mddefault}{\updefault}not allowed!}}}
\put(5776,689){\line( 1, 0){600}}
\put(5776,239){\line( 1, 0){600}}
\put(6676,689){\line( 1, 0){600}}
\put(6676,239){\line( 1, 0){600}}
\put(7576,689){\line( 1, 0){600}}
\put(7576,239){\line( 1, 0){600}}
\put(6976,239){\vector( 0, 1){450}}
\put(4876,689){\line( 1, 0){600}}
\put(8551,689){\line( 1, 0){600}}
\put(4426,614){\makebox(0,0)[b]{\smash{\SetFigFont{10}{12}{\rmdefault}{\mddefault}{\updefault}$D_{5/2}$}}}
\put(4426,239){\makebox(0,0)[b]{\smash{\SetFigFont{10}{12}{\rmdefault}{\mddefault}{\updefault}$P_{3/2}$}}}
\put(5926, 14){\makebox(0,0)[lb]{\smash{\SetFigFont{10}{12}{\rmdefault}{\mddefault}{\updefault}$w_{-1}$}}}
\put(6826, 14){\makebox(0,0)[lb]{\smash{\SetFigFont{10}{12}{\rmdefault}{\mddefault}{\updefault}$w_0$}}}
\put(7726, 14){\makebox(0,0)[lb]{\smash{\SetFigFont{10}{12}{\rmdefault}{\mddefault}{\updefault}$w_{+1}$}}}
\put(6976,-361){\makebox(0,0)[b]{\smash{\SetFigFont{10}{12}{\rmdefault}{\mddefault}{\updefault}allowed!}}}
\end{picture}
\caption{Selective excitation of sublevel populations: not allowed for a transition between
$P_{3/2}$ states but possible for a transition between a $P_{3/2}$ and a $D_{5/2}$ state,
according to degree of controllability analysis.}\label{fig2}
\end{center}
\end{figure}
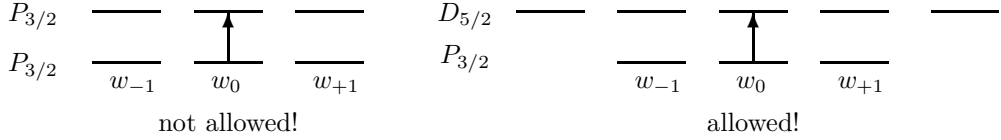

\begin{example}
Suppose we have a system whose ground state is a $P_{3/2}$ state with three sublevels.
Assume that the initial populations of these sublevels are $w_{-1}$, $w_0$ and $w_{+1}$, 
where $w_{\pm 1}\neq 0$, $w_0\neq 0$ and $w_{-1}+w_0+w_{+1}=1$.  Consider a transition 
to another (initially unoccupied) $P_{3/2}$ (triplet) state.  In this case, our analysis
shows that it is \emph{impossible} to \emph{selectively} excite the population of any of
the sublevels.  See \ref{appendix:B} for details.  For instance, we cannot selectively 
transfer the population of the $m=0$ sublevel to the $m=0$ level on the upper electronic
surface without disturbing the populations of the other sublevels.  This may not seem 
surprising as it would be difficult to imagine a pulse scheme that would accomplish such
a task.  However, if in the previous example the upper level is a state with \emph{more}
than three sublevels (e.g., a $D_{5/2}$ or $F_{7/2}$ state) then selective excitation of
\emph{any} one of the sublevels \emph{is} possible since, in principle,  we can generate
any unitary operator, including the one corresponding to selective excitation of any 
sublevel.  See figure \ref{fig2}.
\end{example}

%%%%%%%%%%%%%%%%%%%%%%%%%%%%%%%%%%%%%%%%%%%%%%%%%%%%%%%%%%%%%%
\subsection{Effect of system modifications on controllability}
%%%%%%%%%%%%%%%%%%%%%%%%%%%%%%%%%%%%%%%%%%%%%%%%%%%%%%%%%%%%%%

\begin{figure}
\begin{center}
\setlength{\unitlength}{3158sp}%
\begingroup\makeatletter\ifx\SetFigFont\undefined%
\gdef\SetFigFont#1#2#3#4#5{%
  \reset@font\fontsize{#1}{#2pt}%
  \fontfamily{#3}\fontseries{#4}\fontshape{#5}%
  \selectfont}%
\fi\endgroup%
\begin{picture}(5144,1380)(579,-586)
\thicklines
\put(751,-361){\vector( 0,-1){  0}}
\put(751,-361){\vector( 0, 1){600}}
\put(1951,-361){\vector( 0,-1){  0}}
\put(1951,-361){\vector( 0, 1){600}}
\put(751,239){\vector(-1, 1){  0}}
\put(751,239){\vector( 1,-1){600}}
\put(1351,239){\vector(-1, 1){  0}}
\put(1351,239){\vector( 1,-1){600}}
\put(1351,239){\vector( 1, 1){  0}}
\put(1351,239){\vector(-1,-1){600}}
\put(1951,239){\vector( 1, 1){  0}}
\put(1951,239){\vector(-1,-1){600}}
\put(1801,239){\line( 1, 0){300}}
\put(601,239){\line( 1, 0){300}}
\put(1201,239){\line( 1, 0){300}}
\put(1201,-361){\line( 1, 0){300}}
\put(601,-361){\line( 1, 0){300}}
\put(1801,-361){\line( 1, 0){300}}
\put(1276,-586){\makebox(0,0)[lb]{\smash{\SetFigFont{10}{12.0}{\familydefault}{\mddefault}{\updefault}$|3\rangle$}}}
\put(1276,314){\makebox(0,0)[lb]{\smash{\SetFigFont{10}{12.0}{\familydefault}{\mddefault}{\updefault}$|4\rangle$}}}
\put(676,314){\makebox(0,0)[lb]{\smash{\SetFigFont{10}{12.0}{\familydefault}{\mddefault}{\updefault}$|2\rangle$}}}
\put(1876,-586){\makebox(0,0)[lb]{\smash{\SetFigFont{10}{12.0}{\familydefault}{\mddefault}{\updefault}$|5\rangle$}}}
\put(1876,314){\makebox(0,0)[lb]{\smash{\SetFigFont{10}{12.0}{\familydefault}{\mddefault}{\updefault}$|6\rangle$}}}
\put(751,-586){\makebox(0,0)[b]{\smash{\SetFigFont{10}{12.0}{\familydefault}{\mddefault}{\updefault}$|1\rangle$}}}
\put(3151,-361){\vector( 0,-1){  0}}
\put(3151,-361){\vector( 0, 1){600}}
\put(3751,-361){\vector( 0,-1){  0}}
\put(3751,-361){\vector( 0, 1){600}}
\put(4951,-361){\vector( 0,-1){  0}}
\put(4951,-361){\vector( 0, 1){600}}
\put(5551,-361){\vector( 0,-1){  0}}
\put(5551,-361){\vector( 0, 1){600}}
\put(3151,239){\vector(-1, 1){  0}}
\put(3151,239){\vector( 1,-1){600}}
\put(3751,239){\vector(-1, 1){  0}}
\put(3751,239){\vector( 1,-1){600}}
\put(4351,239){\vector(-1, 1){  0}}
\put(4351,239){\vector( 1,-1){600}}
\put(4951,239){\vector(-1, 1){  0}}
\put(4951,239){\vector( 1,-1){600}}
\put(3751,239){\vector( 1, 1){  0}}
\put(3751,239){\vector(-1,-1){600}}
\put(4351,239){\vector( 1, 1){  0}}
\put(4351,239){\vector(-1,-1){600}}
\put(4951,239){\vector( 1, 1){  0}}
\put(4951,239){\vector(-1,-1){600}}
\put(5551,239){\vector( 1, 1){  0}}
\put(5551,239){\vector(-1,-1){600}}
\put(4801,239){\line( 1, 0){300}}
\put(3601,239){\line( 1, 0){300}}
\put(4201,239){\line( 1, 0){300}}
\put(4201,-361){\line( 1, 0){300}}
\put(3601,-361){\line( 1, 0){300}}
\put(4801,-361){\line( 1, 0){300}}
\put(3001,239){\line( 1, 0){300}}
\put(5401,239){\line( 1, 0){300}}
\put(3001,-361){\line( 1, 0){300}}
\put(5401,-361){\line( 1, 0){300}}
\put(3076,-586){\makebox(0,0)[lb]{\smash{\SetFigFont{10}{12.0}{\familydefault}{\mddefault}{\updefault}$|1\rangle$}}}
\put(3076,314){\makebox(0,0)[lb]{\smash{\SetFigFont{10}{12.0}{\familydefault}{\mddefault}{\updefault}$|2\rangle$}}}
\put(3676,-586){\makebox(0,0)[lb]{\smash{\SetFigFont{10}{12.0}{\familydefault}{\mddefault}{\updefault}$|3\rangle$}}}
\put(3676,314){\makebox(0,0)[lb]{\smash{\SetFigFont{10}{12.0}{\familydefault}{\mddefault}{\updefault}$|4\rangle$}}}
\put(4201,-586){\makebox(0,0)[lb]{\smash{\SetFigFont{10}{12.0}{\familydefault}{\mddefault}{\updefault}$|5\rangle$}}}
\put(4276,314){\makebox(0,0)[lb]{\smash{\SetFigFont{10}{12.0}{\familydefault}{\mddefault}{\updefault}$|6\rangle$}}}
\put(4876,-586){\makebox(0,0)[lb]{\smash{\SetFigFont{10}{12.0}{\familydefault}{\mddefault}{\updefault}$|7\rangle$}}}
\put(4876,314){\makebox(0,0)[lb]{\smash{\SetFigFont{10}{12.0}{\familydefault}{\mddefault}{\updefault}$|8\rangle$}}}
\put(5476,-586){\makebox(0,0)[lb]{\smash{\SetFigFont{10}{12.0}{\familydefault}{\mddefault}{\updefault}$|9\rangle$}}}
\put(5401,314){\makebox(0,0)[lb]{\smash{\SetFigFont{10}{12.0}{\familydefault}{\mddefault}{\updefault}$|10\rangle$}}}
\put(1501,614){\makebox(0,0)[b]{\smash{\SetFigFont{10}{12.0}{\rmdefault}{\bfdefault}{\updefault}Ex 3c(2): F=1 / F=1}}}
\put(4426,614){\makebox(0,0)[b]{\smash{\SetFigFont{10}{12.0}{\rmdefault}{\bfdefault}{\updefault}Ex 5c(2): F=2 / F=2}}}
\end{picture}
\caption{Coupling diagrams for transitions between two $P_{3/2}$ and $D_{5/2}$ states,
         respectively, for certain alkali atoms.  The transition between $m=0$ states 
         is prohibited even for linearly polarized light.  Nevertheless, controllability
         analysis shows that the degree of controllability of the system remains the
         same.}\label{fig3}
\end{center}
\end{figure}
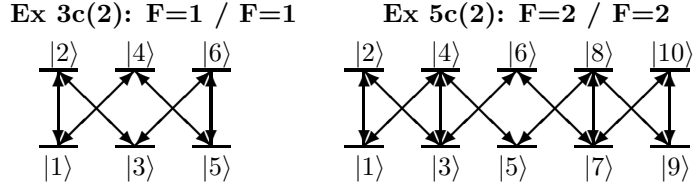

The question also arises, how small modifications of the system affect the degree of
controllability.  For instance, for alkali atoms transitions between the $m=0$ sublevels
are prohibited for transitions between two states with the same $F$ values, even for 
linearly polarized light.  Thus, given a transition between two $P_{3/2}$ or $D_{5/2}$ 
states, for example, we must modify the transition diagrams in examples 3a and 5a as shown
in figure \ref{fig3}.  To reflect this change in the model, the values of the corresponding
dipole moments in the Hamiltonian $\op{H}_1$ (see equations (\ref{H1:F=1}) and 
(\ref{H1:F=2}), respectively) are set equal to zero.  Nevertheless, computation of 
the Lie algebra shows that in both cases the Lie algebra remains isomorphic to $sp(3)$ 
and $sp(5)$ respectively, even with the same $\op{J}$ matrix.  Thus, the changes do not
affect the degree of controllability of the system.

%%%%%%%%%%%%%%%%%%%%%%%%%%%%%%%%%%%%%%%%%%%%%%%%%%%%%%%%%%%%%%
\section{Limits of controllability analysis and open problems}
%%%%%%%%%%%%%%%%%%%%%%%%%%%%%%%%%%%%%%%%%%%%%%%%%%%%%%%%%%%%%%

%%%%%%%%%%%%%%%%%%%%%%%%%%%%%%%%%%%%%%%%%%%%%%%%%%%%%%%%%%%%%%%%%
\subsection{Constructive control using the Lie algebra structure}
%%%%%%%%%%%%%%%%%%%%%%%%%%%%%%%%%%%%%%%%%%%%%%%%%%%%%%%%%%%%%%%%%

The degree of controllability of a system allows us to determine what control objectives
are realistic, at least in principle, which is important for practical applications.  As
we have shown in the previous section, while controllability analysis based on the dynamical 
Lie algebra of the system often confirms our intuition about the controllability of the
system, it can sometimes lead to surprising results, such as the difference in the degree
of controllability for transitions between states with the same $F$ value, and those 
between states with different $F$ values, etc.

Precise knowledge of the dynamical Lie algebra of the system can also be used to establish
(explicitly verifiable) criteria for dynamical realizability of unitary operators, as well
as reachability/non-reachability of quantum states \cite{qph0110171, IEEE40CDC1122}.  
However, in general, controllability analysis does not provide a constructive procedure for
realizing a desired control objective that has been shown to be theoretically achievable.
Many techniques have been proposed to solve the challenging problem of finding suitable 
control fields.  Most rely on some form of numerical optimization, model-based feedback or
learning algorithms \cite{PRA61n012101, JCP110p9825, JCP109p385, ACP101p315, JPA33p4643, 
JCP109p9318, PRA62n012105, PRA62n012307, PRA60p2700, JCP110p7142}.

The problem of constructive control using the Lie algebra structure of the system has also
been addressed for certain systems such as coupled spin systems, where \emph{explicit} 
procedures for generating arbitrary unitary operators have been developed \cite{CP267p11}.
Constructive control techniques based on Lie group decompositions have also been developed
to derive control schemes for explicit generation of unitary operators \cite{PRA61n032106,
PRA62n053409,CP267p25,IEEE39CDC1074,dalessandro}, and to achieve other control objectives 
such as maximization of observables and quantum state engineering \cite{qph0105155} in 
$N$-level quantum systems with selectively addressable transitions.  In \cite{IEEE40CDC0306}
similar techniques have been used to construct control sequences for a few target operators
for some simple systems with degenerate transition frequencies and symmetrically coupled 
transitions.  However, as regards \emph{constructive} controllability for atomic systems 
with degenerate energy levels and transitions, further work is necessary.

%%%%%%%%%%%%%%%%%%%%%%%%%%%%%%%%%%%%%%%%%%%%
\subsection{Controllability of open systems}
%%%%%%%%%%%%%%%%%%%%%%%%%%%%%%%%%%%%%%%%%%%%

In this paper we systematically studied the degree of controllability of closed quantum 
control systems, i.e., quantum systems that only interact with a set of external control
fields, whose interaction Hamiltonian is Hermitian.  The question of controllability of
open quantum systems, i.e., systems that also interact with the environment, is rather 
more complicated to answer.  The time evolution operator of an open system depends 
critically on the interaction with the bath, represented by a non-Hermitian operator, and
the dynamical evolution of the system is thus \emph{not} unitary.  Therefore, there are no
conservation laws for energy, entropy and probability for such systems. 

This lack of unitary evolution and conservation laws for open quantum systems can be an 
advantage or an impediment.  For instance, dissipation, e.g., by spontaneous emission, 
is \emph{essential} for important applications such as optical pumping, quantum reservoir
engineering with laser cooled trapped ions \cite{PRL77p4728}, and laser cooling of internal
molecular degrees of freedom \cite{PRA63n013407,FD113p365,JCP106p1435}.  Thus, one might 
say that dissipation enhances the controllability of these systems.  On the other hand, 
there are applications such as quantum information processing, where dissipative effects
such as decoherence are extremely deleterious.  This suggests that the effect of dissipation
on the degree of controllability of open quantum control systems should be assessed on a 
case-by-case basis.  

%%%%%%%%%%%%%%%%%%%%
\section{Conclusion}
%%%%%%%%%%%%%%%%%%%%

We have provided precise definitions for various degrees of controllability for closed
quantum control systems as well as necessary and sufficient conditions for each, and 
applied the theoretical results to determine the degree of controllability of various 
atomic systems with degenerate energy levels and transition frequencies.  Although the 
results of this controllability analysis based on the dynamical Lie algebra of the system
often confirm our intuition about the controllability of the system, they can sometimes
be surprising, such as the difference in the degree of controllability for transitions 
between states with the same degree of degeneracy, and those between states with different
$F$-values.  

At present, our controllability analysis for atomic systems with degenerate energy levels
and transition frequencies allows us only to determine the \emph{degree of controllability}
of the system, which can then be used to assess the \emph{feasibility of certain control 
objectives}.  In the future, it may be possible to use the structure of the Lie algebra 
of these systems to develop \emph{constructive techniques} for the generation of unitary 
operators using simple control pulse sequences, as has been done for coupled spin systems
and $N$-level systems with selectively addressable transitions.

\section*{Acknowledgements}

SGS acknowledges the hospitality and financial support of the University of Oregon and 
thanks the faculty of the Department of Mathematics and Institute of Theoretical Science
at the University of Oregon for many valuable discussions.  AIS acknowledges the 
hospitality of the Laboratoire de Physique Th\'eorique des Liquides, University of Paris
VI.  The authors also thank V. Ramakrishna and G.\ Turinici for valuable discussions.

\section*{References}
\bibliography{books,papers2000,papers9599,papers9094,papers--79}

\appendix
\section{Identifying the dynamical Lie algebra}
\label{appendix:A}

In order to decide whether the dynamical Lie algebra of a system with Hamiltonian
(\ref{eq:H}) is of type $sp(\frac{N}{2})$, $su(N)$ or $u(N)$ we proceed as follows.  
We first compute the dimension of the Lie algebra generated by the operators $\rmi
\tilde{H}_m$, i.e., the trace-zero parts of the $\rmi\op{H}_m$ using the algorithm
described in \cite{PRA63n063410}.  If the dimension of the Lie algebra $\tilde{\L}$
is $N^2-1$ we check the trace of the operators $\op{H}_m$.  If at least one of the
operators has non-zero trace then the Lie algebra is isomorphic to $u(N)$ and the
system is completely controllable.  If the dimension of $\tilde{\L}$ is $N^2-1$ and
all $\op{H}_m$ have zero trace then $\L=su(N)$ and the system is both density matrix
and observable controllable.   If $N$ is even and $\tilde{\L}$ has dimension $N(N+1)/2$
then we check if there exists an operator $\tilde{J}$, which is unitarily equivalent
to the $\op{J}$ defined in equation (\ref{eq:Jsp}) such that
\begin{equation} \label{eq:sp-cond}
  (\rmi\tilde{H}_m)^T\tilde{J}+\tilde{J}(\rmi\tilde{H}_m) =0
\end{equation}
holds for all $m$.  If such a $\tilde{J}$ exists then we conclude that the Lie algebra 
$\tilde{\L}$ is isomorphic to a subalgebra of $sp(\frac{N}{2})$ and as the dimension of
the subalgebra equals the dimension of $sp(\frac{N}{2})$ it follows that $\tilde{\L}
\simeq sp(\frac{N}{2})$.

To determine if there exists a $\tilde{J}$ such that (\ref{eq:sp-cond}) is satisfied
we note that (\ref{eq:sp-cond}) can be rewritten as system of linear equations 
\[
  \tilde{\L}_m \vec{J} = 0, \quad 0\le m\le M,
\]
where $\tilde{\L}_m$ is an $N^2\times N^2$ matrix determined by the $\rmi\tilde{H}_m$ 
and $\vec{J}$ is an $N^2$ column vector.  To determine the solutions $\vec{J}$ of the 
above equation we compute the null space of 
\[
  \left( \begin{array}{c} \tilde{\L}_0 \\ \vdots \\ \tilde{\L}_M \end{array} \right).
\]
If it is non-empty then there is a $\tilde{J}$ such that (\ref{eq:sp-cond}) is satisfied
and we can compute its eigenvalues and compare them to the eigenvalues of the standard 
$\op{J}$ given in (\ref{eq:Jsp}), which allows us to determine if this $\tilde{J}$ is 
unitarily equivalent to the standard $\op{J}$.  (Recall that two operators are unitarily
equivalent if and only if the have the same eigenvalues.)

\section{Identifying forbidden operations}
\label{appendix:B}
To show that selective excitation of the $m=0$ sublevel population for a transition 
between two $P_{3/2}$ states is indeed impossible in the context of this model, no 
matter what fields are employed, we note that this selective excitation would require
implementing a unitary operator $\op{U}$ that permutes the populations of states 
$\ket{3}$ and $\ket{4}$ (see Example 3a for notation) without disturbing the other
states.  Such a unitary operator would have to be of the form
\[
  \op{U} = \left(\begin{array}{cc|cc|cc} 1 & 0 & 0 & 0 & 0 & 0 \\
                                         0 & 1 & 0 & 0 & 0 & 0 \\\hline
                                         0 & 0 & 0 & 1 & 0 & 0 \\
                                         0 & 0 & 1 & 0 & 0 & 0 \\\hline
                                         0 & 0 & 0 & 0 & 1 & 0 \\
                                         0 & 0 & 0 & 0 & 0 & 1
                  \end{array}\right)
\]
modulo phase factors for the non-zero entries.  However, it can easily be verified 
using the technique outlined in the previous section, that the Lie algebra generated
by the system, i.e., by $\rmi\op{H}_m$ with $\op{H}_m$ as in equations (\ref{H0:F=1}),
(\ref{H1:F=1}) and (\ref{H2,3:F=1}), is isomorphic to $sp(3)$ with 
\[
   \op{J}= \left(\begin{array}{cccccc} 0 & 0 & 0 & 0 & 0 & +1 \\
                                       0 & 0 & 0 & 0 & -1 & 0 \\
                                       0 & 0 & 0 & +1 & 0 & 0 \\
                                       0 & 0 & -1 & 0 & 0 & 0 \\
                                       0 & +1 & 0 & 0 & 0 & 0 \\
                                       -1 & 0 & 0 & 0 & 0 & 0
                  \end{array}\right).
\]
Any unitary operator generated by this system must preserve this $\op{J}$, i.e.,
$\op{U}^T\op{J}\op{U}=\op{J}$.  It is easy to verify that the $\op{U}$ above does
not preserve $\op{J}$.  Hence, it \emph{cannot} be implemented.  It is similarly
easy to show that selective excitation of the $m=-1$ or $m=+1$ sublevel is not
possible for this system.
\end{document}